\documentclass[prl,twocolumn,superscriptaddress,aps]{revtex4}
\usepackage{amssymb,graphics,graphicx}
\usepackage{latexsym}
\usepackage{amsfonts}
\usepackage{amsmath}
\newcommand{\be}{\begin{equation}} \newcommand{\ee}{\end{equation}}

\begin{document}

\title{Conformal Completion of the Standard Model with a Fourth Generation}
\author{Chiu Man Ho}
\email{chiuman.ho@vanderbilt.edu}
\affiliation{Department of
  Physics and Astronomy, Vanderbilt University, Nashville, TN 37235, USA}
\author{Pham Q. Hung}
\email{pqh@virginia.edu}
\affiliation{Department of Physics, University of Virginia,
Charlottesville, VA 22904-4714, USA}
\author{Thomas W. Kephart}
\email{tom.kephart@gmail.com}
\affiliation{Department of Physics and Astronomy, Vanderbilt University, Nashville, TN 37235, USA}

\date{\today}

\begin{abstract}

We study dynamical electroweak symmetry breaking with a fourth generation within
the $Z_n$ orbifolded $AdS_5\otimes S^5$ framework.
A realistic  $Z_7$ example is discussed. The initial theory reduces dynamically, due to the induced condensates, to a four-family
trinification near a TeV-scale conformal fixed point
where the gauge hierarchy problem does not exist.
We predict new gauge bosons and bifundamental fermions and scalars accessible by the LHC.

\end{abstract}

\pacs{}\maketitle

\section{Introduction}

The fact that the renormalization group running of mass parameters spoils any fine tuning between the grand unification scale and the electroweak scale is the long-standing hierarchy problem of the standard model (SM) of particle physics. There have been a number of suggestions of how to ameliorate or even solve this problem, but most do not come with testable predictions. Supersymmetry (SUSY) is one example that lessens the problem and has predictions, but the overhead of doubling the particle spectrum is somewhat unattractive. Another approach to stabilizing running is via conformality. Conformal invariance implies scale invariance, hence the hierarchy problem should not arise in a conformal model, since there are no masses to run. Once conformal symmetry is broken and mass terms appear and they can run, but if the breaking scale is near the electroweak scale, then the fine tuning required for a hierarchy is not needed. Hence, it is important that we search for extensions of the standard model that are either conformal or that can evolve to conformal fixed points. Just such a model has recently been discussed \cite{Hung} where a fourth generation fermion is introduced to provide the stabilization \cite{Frampton:1999xi}. What remains to be shown is how such a model can fit into a more fundamental ultraviolet completion (where ultraviolet completion must be discussed with caution, since once the model is conformal, it is in a sense already
UV complete).

From the string theory side, i.e., from an ultra high energy perspective, we know a large class of conformal or potentially conformal models.
These are models based on the AdS/CFT correspondence. To be more specific, we will only consider $Z_n$ orbifolded $AdS_5\otimes S^5$ models
\cite{Frampton:2007fr,Aldazabal:2000sa}, but the class of potential theories is much wider. What we will seek is a matching between a four-family low-energy model
with conformal unifcation scale in the TeV region, and an orbifolded model from the string scale.


\section{Phenomenology of a Fourth Generation}

Recently, the idea of a possible fourth generation has attracted a lot of attention \cite{Hou}.
CDF has searched for the direct production of the 4th generation quarks $t'$ and $b'$,
assuming $t'\rightarrow W \,q $ and $b'\rightarrow W \,t $ \cite{CDF,whiteson}. This places the lower mass bounds $m_{t'} > 335$ GeV and $m_{b'} > 385$ GeV, while the upper bound allowed by partial wave unitarity is about 600 GeV \cite{Chanowitz}.
The direct mass limits on the $t'$ and $b'$ quarks in all mixing scenarios have been studied in \cite{TimTait}.
For a critical discussion of the experimental constraints on the 4th generation quark masses, see \cite{Hung:2007ak}.
Even if the $t'$ and $b'$ quarks are too heavy to be seen directly, their effects may be manifest at the LHC since they induce
a large  $gg \rightarrow ZZ$ signal \cite{Chanowitz2}. Assuming the presence of 4th generation fermions with large masses,
the combined analysis from CDF and D0 has excluded a standard-model-like Higgs boson with a mass between 131 GeV and 204 GeV at 95\% confidence
level \cite{D0}. The existence of heavy 4th generation quarks can also enhance both of the processes $gg \rightarrow H $ and $H \rightarrow gg $ \cite{higgsphysics}.
Recently, a four-family MSSM has been explored \cite{Dawson}. Imposing perturbative unitarity and constraints from precision electroweak
data, it was shown that the four-family MSSM is consistent only with $\tan \beta \sim 1$ and highly tuned fermion masses. This seems to
suggest that SUSY and 4th generation fermions may not be a good fit. In fact, as we have mentioned above,
a 4th generation with a conformal fixed point solves the hierarchy problem naturally and supersymmetry is completely unnecessary. A fourth generation is also compatible with light fermion flavor symmetry~\cite{Chen:2010ty}.    Thus, in this Letter, we will search for non-SUSY conformal four-family models from orbifolded $AdS_5\otimes S^5$.


\section{Orbifolded $AdS_5\otimes S^5$ Models}

Given a orbifolding group $\Gamma$,
we will base our choices of suitable  $AdS_5\otimes S^5/\Gamma$ models on the following criteria. First, we look for models which yield four chiral families at the end of a chain of symmetry breaking. Second, the conformal scale should not exceed $\sim 10 \,$ TeV. This requirement comes from the fact that the conformal scale is related to the mass scale of the 4th generation which cannot be too ``low" experimentally: The higher the conformal scale is, the lower the mass scale of the 4th generation becomes as shown in \cite{Hung}. Third, the model should yield the correct value of $\sin ^2 \theta_W(M_Z)\approx 0.231$. This constrains the initial value  $\sin ^2 \theta_W^0$ at the conformal scale (assumed to be of O($1-10 \,$ TeV)) to be neither too low nor too high as we shall see below. Fourth, $\alpha_3(M_Z)$ should be in agreement with experiment, namely $\alpha_3(M_Z)\approx 0.118$.

As is well known, the original $SU(N)$ conformal $AdS_5\otimes S^5$
theory \cite{footnote} has ${\cal N}=4$ SUSY. With the replacement of $S_5$ by an orbifold $S_5/\Gamma$, the ${\cal N}=4$ SUSY can be broken down to ${\cal N}=2,1,0$ if  $\Gamma$ is embedded nontrivially in $SU(2),SU(3)$ and $SU(4)$ respectively.  In what follows, we shall work with a non-supersymmetric scenario ${\cal N}=0$ where $\Gamma$ is a discrete subgroup of $SU(4)$, chosen for simplicity to be $\Gamma=Z_n$.
In this breaking of ${\cal N}=4$ down to ${\cal N}=0$, the resultant gauge group is $SU(N)^n$. Furthermore, we wish to embed the SM into this resultant gauge group. As a result, we shall choose $N=3$, and then be searching for the appropriate $n$ which yields four standard model generation. While the initial $SU(N)$ conformal $AdS_5\otimes S^5$ is assumed to be conformally invariant in the large $N$ limit, the related finite $N$ orbifolded models in general are not.
However, they are one-loop finite and some have been shown to be two-loop finite \cite{Fuchs}. So the orbifolded variants have an approximate conformal symmetry to some degree. The breaking of conformal invariance, either due to orbifolding or finite $N$, introduces a mass scale, call it $\Lambda$. At the phenomenological level, this scale will propagate into effective terms in the Lagrangian. In what follows we will assume $\Lambda$ is introduced through dimension five terms of the form $\psi^2\phi^2/\Lambda$.

For the non-supersymmetric scenario ${\cal N}=0$ with $\Gamma = Z_n$, the four-dimensional representation of $SU(4)$ is
given by ${\bf  R_4}=(\alpha^{n_1},\alpha^{n_2},\alpha^{n_3},\alpha^{n_4})$ where \,$\alpha = \exp\,(\frac{2\,\pi\,i}{n})$\, and\,
$n_1+n_2+n_3+n_4 =0$ \,mod\, $n$. As pointed out by \cite{weakHooft}, there exist twisted operators $O_l$ that transform by $\exp\,(\frac{2\,l\,\pi\,i}{n})$ under the generators of $Z_n$. In the weak 't Hooft coupling limit, it has been shown that double-trace operators of the form $ O\, \bar{O}$ may be induced in the effective action \cite{weakHooft}. These operators render the gauge theory non-conformal because the one-loop beta functions of the corresponding double-trace couplings do not possess real zeros. The existence of these double-trace operators requires $O$ and $\bar{O}$ to have opposite quantum numbers under $\Gamma = Z_n$, namely $ O_{l}\, \bar{O}_{-l}$ for $\Gamma = Z_n$. This requirement translates into the conditions \, $n_1=-n_2 = n'$ \,(mod $n$)\, and \, $n_3=-n_4 = n''$ \,(mod $n$). These conditions are indicative of vectorlike models. However, neither the toy $Z_4$ model, nor the more realistic $Z_7$ model to be presented in the paper, both of which are chiral models,  satisfy these relations required by the double trace operator.  As a result, the double-trace operators, which spoil  conformal symmetry, do not exist in our models. In the strong 't Hooft coupling limit, it has been shown that there are non-perturbative instabilities associated the $AdS_5\otimes S^5/Z_n$ theories that break all the supersymmetry \cite{strongHooft}. However, these non-perturbative instability would not occur if we assume that we are working in the weak 't Hooft limit.

In the search for an acceptable conformal model, we will rely on the way the SM is embedded into a $SU(3)^3$ diagonal subgroup of  $SU(3)^n$. For instance, if we write $SU(3)^n = SU(3)^{n-p-q}\times SU(3)^p\times SU(3)^q\rightarrow SU(3)_c \times SU(3)_L \times SU(3)_R$, where each subgroup in the product has broken to   its diagonal $SU(3)$ subgroup, then the gauge couplings need to be renormalized for each diagonal subgroup to preserve the commutation relations of the generators. The general result for $SU(3)^k\rightarrow SU_D(3)$ is $g\rightarrow g/\sqrt{k}$ and hence $\alpha \rightarrow \alpha/k$. Now from the  trinification group $SU(3)_c \times SU(3)_L \times SU(3)_R$, where $SU(2)_L  \subset SU(3)_L$ , with the hypercharge $Y$ distributed as usual, two thirds in $SU(3)_L$ and one third in $SU(3)_R$, we can compute the formula for the weak mixing angle \cite{Kephart:2004qp,Frampton:2007fr,HBB}:
\begin{equation}
\sin ^2 \theta_W^0 = \frac{3}{3 +5 \left(\frac{p+2q}{3p}\right)}\,.
\end{equation}

A compelling model is one which can satisfy all of the aforementioned criteria. We search for models with four families at the trinification scale, i.e.,
\begin{equation}
\label{trinification}
4[(3,\bar3,1)+(1,3,\bar3)+(\bar3,1,3)]_F\,.
\end{equation}
For clarity, it is helpful to explicitly list the particle content for each family given in (\ref{trinification}). For this purpose, we will use the notation of \cite{Leser:2011fz}, namely
\begin{eqnarray}
\label{content1}
\psi_L=(1,3,\bar3)=\left(
\begin{array}{ccc}
    {\cal E}_L     & E^c_L      & {\cal L}_L       \\
    {\cal N}_{1L} & e^{+}_{L} & {\cal N}_{2L} \\
  \end{array}
\right)  \,,
\end{eqnarray}
\begin{eqnarray}
\label{content2}
\psi_{Q^c}=(\bar3,1,3)=\left(
        {\cal D}^c,
        u^c,
        {\cal B}^c
\right)_L  \,,
\end{eqnarray}
\begin{eqnarray}
\label{content3}
\psi_Q=(3,\bar3,1)=\left(
  \begin{array}{c}
    \left(
      \begin{array}{c}
        -d \\
        u \\
      \end{array}
    \right)_L
     \\
     B_L\\
  \end{array}
\right) \,,
\end{eqnarray}
where
\begin{eqnarray}
{\cal E}_L=\left(
      \begin{array}{c}
        \nu_1 \\
        e_1 \\
      \end{array}
    \right)_L ~;~ E^c_L=\left(
         \begin{array}{c}
            E^+\\
            N^c \\
         \end{array}
       \right)_L ~;~ {\cal L}_L=\left(
          \begin{array}{c}
            \nu_2 \\
            e_2 \\
          \end{array}
        \right)_L\,.
\end{eqnarray}
As reviewed in \cite{Leser:2011fz}, one combination of ${\cal E}_L$ and ${\cal L}_L$ gives rise to one SM lepton doublet while the orthogonal combination will become a new heavy lepton doublet. In addition, there is a non-standard doublet $E^c_L$. For the quarks, one combination of ${\cal D}^c$ and ${\cal B}^c $ gives the right-handed down quark while the orthogonal combination gives a new charge -1/3 heavy quark. We shall come back to this particle content below.


The simplest irreducible non-SUSY four-family
model is an orbifolded $AdS_5\otimes S^5/Z_4$ with $Z_4$ irreducibly embedded in the four-dimensional representation of $SU(4)$ as ${\bf  R_4}=(\alpha,\alpha,\alpha,\alpha)$. The resulting ${\cal N}=0 $ $SU(3)^4$ quiver gauge theory has the fermions in the bifundamental representation: \,4\,$\sum_{i=1}^{4~\textrm{mod}[4]}\,(\,3_i,\,\bar3_{i+1}\,)$. These fermions transform as
$4[(3,\bar3,1,1)+(1,3,\bar3,1)+(1,1,3,\bar3)+(\bar3,1,1,3)]_F\,.$
The scalars of the theory are determined by the corresponding embedding in the
${\bf R_6}=(\alpha^2,\alpha^2,\alpha^2,\alpha^2,\alpha^2,\alpha^2)$ and are also in the
bifundamental representation: \,3\,$\sum_{i=1}^{4~\textrm{mod}[4]}\,(\,3_i,\,\bar3_{i\pm 2}\,)$. These scalars transform as
$6[(3,1,\bar3,1)+(1,3,1, \bar3)+(\bar3,1,3,1)+(1,\bar3,1,3)]_S\,.$

In the breaking of $SU(3)^4$ down to $SU(3)^3$, in order for the fermion content to reduce to four chiral families (\ref{trinification}), one needs to have a condensate of the form $\langle (1,1,3,\bar{3}) \rangle$. But the scalar sector is insufficient since it does not have any field carrying the desired quantum numbers.
Another shortcoming of the $Z_4$ model is that all four $SU(3)$ gauge couplings start off equal, and even if we could break the symmetry, the gauge couplings could not be sufficiently split.
We take this point as the motivation to explore more complicated orbifolding.

In our search, we have found the following $Z_n$ models which yield four chiral families and values of  $\sin ^2 \theta_W^0$ which lie in the desired range: \\
(I)\; ~~~$Z_7$ ($p=2$, $q=4$) with $\sin ^2 \theta_W^0=0.265$\,;\\
(II)\, ~~$Z_{11}$ ($p=3$, $q=7$) with $\sin ^2 \theta_W^0=0.241$\,; \\
(III)\;~ $Z_{13}$ ($p=4$, $q=8$) with $\sin ^2 \theta_W^0=0.265$\,;\\
(IV)\; ~$Z_{14}$ ($p=4$, $q=9$) with $\sin ^2 \theta_W^0=0.247$\,.\\

Let us focus on the $\Gamma= Z_7$ model which is the first case where the problems of symmetry breaking and gauge coupling degeneracy can be avoided \cite{Frampton:1999wz}.
We will discuss how $SU(3)^7$ can dynamically break to $SU(3)^3$ through fermion condensates, resulting in four massless chiral families and TeV-scale exotic fermions before electroweak symmetry breaking.

We consider an orbifolded $AdS_5\otimes S^5/Z_7$ where the embedding is  ${\bf  R_4}=(\alpha,\alpha^2,\alpha^2,\alpha^2)$ and the gauge group is $SU(3)^7$. The fermions are in the bifundamental representation: \,$\sum_{i=1}^{7~\textrm{mod}[7]}\,[\,(\,3_i,\,\bar3_{i+1}\,)+3\,(\,3_i,\,\bar3_{i+2}\,)\,]$.
The scalars are derived from the embedding ${\bf  R_6}=(\alpha^3,\alpha^3,\alpha^3,\alpha^4,\alpha^4,\alpha^4)$ and are also in the
bifundamental representation: \,3\,$\sum_{i=1}^{7~\textrm{mod}[7]}\,(\,3_i,\,\bar3_{i\pm 3}\,)$.
As we shall see,
these fermions and scalars provide enough degrees of freedom to break $SU(3)^7$  to $SU(3)^3$, but first we take a detour to describe the mechanism for dynamical symmetry \cite{Hung2}, since it is relevant to the aforementioned symmetry breaking.

\section{A Brief Review of Dynamical Electroweak Symmetry Breaking with a Heavy Fourth Generation}

At the two-loop level, the Yukawa couplings of a heavy 4th generation reach a quasi fixed point at a scale $\Lambda_{FP} \sim O(\textrm{TeV})$ \cite{Hung}. (It is at $O(10^{16}\,\textrm{GeV})$ in the unrealistic case of a light 4th generation.) This study hints at the possibility that scale invariance is restored at  $\Lambda_{FP}$ and beyond. As  mentioned above,  conformal invariance implies scale invariance and one can envision a scenario, such as the one presented here, in which a four generation standard model (SM4) merges into a conformally invariant model at around  $\Lambda_{FP}$, which was taken to be the physical cut-off scale in \cite{Hung}. Above $\Lambda_{FP}$, all masses are absent. The spontaneous breaking of conformal invariance will generate mass scales and, in particular, those associated with the electroweak masses. The question  explored in \cite{Hung2} was: Could dynamical electroweak symmetry breaking occur with a heavy 4th generation using only the degrees of freedom present in SM4?

A fourth generation which is sufficiently heavy can lead to the formation of bound states and condensates which, in turn, can trigger dynamical electroweak symmetry breaking \cite{Hung,Burdman}. A detailed study of how condensates get formed using the ladder approximation in the Schwinger-Dyson (SD) equation was carried out in \cite{Hung2} where it was found that, starting with a scale-invariant SM4 where the fundamental Higgs field is massless, condensates carrying the electroweak quantum numbers get formed by the Higgs exchange coming from a Yukawa interaction of the type ${\cal L}_Y = - g_{b'} ~\bar{q}_L \Phi ~b'_R - g_{t'} ~\bar{q}_L \widetilde{\Phi} ~t'_R + h.c. $. It was shown that there exists a critical Yukawa coupling above which condensate formation is allowed, namely $\alpha_c \equiv g^2_Y/4\pi = \pi/2 \approx 1.57$. Basically, the SD equation can be summarized as a differential equation
\begin{equation}
\label{SD}
\Box \Sigma_{4Q}(p) = - (\frac{\alpha_{4Q}}{\alpha_c}) \frac{\Sigma_{4Q}(q)}{q^2 + \Sigma_{4Q}^2(q)} \,,
\end{equation}
supplemented by two boundary conditions
\begin{eqnarray} \label{bc1}
&&\lim_{p \to 0} p^4 \frac{d \Sigma_{4Q}}{d p^2} = 0 \,, \nonumber \\
&&\lim_{p \to \Lambda}  p^2 \frac{d \Sigma_{4Q}}{d p^2} + \Sigma_{4Q}(p)  = 0 \,.
\end{eqnarray}
In Eq.(\ref{bc1}), $\Lambda$ represents the {\em energy scale where the boundary condition is obeyed}. For electroweak symmetry breaking,  $\Lambda \sim \Lambda_{FP} \sim O(TeV)$.

These condensates, $\langle \bar{t'}_L t'_R \rangle=\langle \bar{b'}_L b'_R \rangle$ for the 4th generation quarks (with similar expressions for the 4th generation leptons), are required, by the knowledge of the W and Z masses, to be of  ${O(-\Lambda_{EW}^3)}$ where $\Lambda_{EW} \sim 246\,$ GeV. They are shown to depend on the cut-off scale which was taken to be $\sim \Lambda_{FP}$ and have implications for the hierarchy problem. If we denote the Yukawa coupling of the 4th generation quarks by $\alpha_{4Q}$, it turns out that the aforementioned requirement for the condensate translates into a relationship $\frac{\alpha_{4Q}^{cond}}{\alpha_c} \sim 1+ (\frac{\Lambda_{EW}}{\Lambda_{FP}})^2$. A {\em large} $\Lambda_{FP}$ (corresponding to an unrealistic light 4th generation) would necessitate an incredible fine tuning by e.g. 28 decimal points for $\Lambda_{FP} \sim 10^{16}\,$ GeV. On the other hand,  {\em no such fine tuning} is needed when  $\Lambda_{FP} \sim O(\textrm{TeV})$.

The SD analysis of condensates as carried out in \cite{Hung2} is for a fixed Yukawa coupling.  To see how SM4 fits into this scheme of dynamical electroweak symmetry breaking, one can look at the evolution of the Yukawa couplings as had been done in \cite{Hung}. Starting with initial values of the Yukawa couplings at the electroweak scale, which would correspond to a mass if they are multiplied by  $\Lambda_{EW}$, the evolution at one and two loops shows a sharp rise in values and eventually exceeding $\alpha_c$ at a scale close to $\Lambda_{FP}$ \cite{Hung}. At $\alpha_{4Q}^{cond} \approx \alpha_c\, (\,1+ (\frac{\Lambda_{EW}}{\Lambda_{FP}})^2\,)$, electroweak condensates from the 4th generation get formed. This value lies between the critical coupling $\alpha_c \approx 1.57$ and the two-loop quasi fixed point $\alpha^* \approx 4$. It was proposed in \cite{Hung2} that $\alpha_{4Q}^{cond} $ might represent the value of the true fixed point of a scale-invariant theory or, as in this paper, a conformal invariant theory.

If SM4 is replaced by a conformal theory above $\Lambda_{FP}$ and if we assume that the true Yukawa fixed point value is $\sim \alpha_{4Q}^{cond}$, it will remain so above that scale. As stressed in \cite{Hung2}, one {\em does not have} the formation of condensates at energy scales above $\Lambda_{FP}$  because one of the boundary conditions Eq. (\ref{bc1}) of the SD equation is required to be satisfied only at $\Lambda_{FP}$. All spontaneous symmetry breaking occur at that scale and {\em not above}. This important point will be applied to the condensate-induced symmetry breaking of the $Z_7$ model .


\section{Dynamical Symmetry Breaking of $SU(3)^7 \rightarrow SU(3) \times SU(2)_L \times U(1)_Y$ via Fermion Condensates}

As in the previous section, we will assume that the relevant scalars
are massless, or at least light, in the conformal region above O(TeV), and that the Yukawa couplings are large enough for condensates to be formed. At each step in the breaking chain, we will explicitly write down the Yukawa interaction among the relevant fields. In what follows, we will sketch  the salient points of the various steps for each class of fermion bifundamentals and leave the details for a longer version. However, we wish to emphasize, for clarity, that the condensates listed below are assumed to occur at energy scales where the boundary conditions to the SD equation are obeyed. In the chain of symmetry breaking, these scales can be different from one another. Their actual possible values will be presented elsewhere.

First we assume that there are effective composite scalars formed from
combining pairs of fundamental scalars. To see how this works, we assume that conformal symmetry is softly broken.  This breaking is expected since the orbifolding does not in general preserve conformal symmetry. The simplest way to allow conformal symmetry breaking (csb) is by adding a scalar mass term of the form $M_{csb}^2\phi^2$, where $M_{csb}$ is related to $\Lambda_{FP}$ introduced above. This mass term in turn induces dimension 5 operators $M_{csb}^{-1}\psi^2\phi^2$ through loops where the $\phi$'s are fundamental fields, and where we will treat  the product $\phi^2$ as an effective composite field for convenience. More explicitly, for the case at hand we write
\begin{eqnarray}
\phi_1 &=& \frac{1}{M_{csb}}\,[\,(1,1,1,\bar3,1,1,3)\otimes (1,1,3,1,1,1,\bar3) \,]\,, \\
\phi_2 &=& \frac{1}{M_{csb}}\,[\,(1,1,\bar3,1,1,3,1)\otimes (1,3,1,1,1,\bar3,1) \,]\,, \\
\phi_3 &=& \frac{1}{M_{csb}}\,[\,(1,\bar3,1,1,3,1,1)\otimes (3,1,1,1,\bar3,1,1) \,]\,, \\
\phi_4 &=& \frac{1}{M_{csb}}\,[\,(1,1,3,1,1,1,\bar3)\otimes (1,1,\bar3,1,1,3,1) \,]\,,
\end{eqnarray}
for the effective operators that we need below. Condensates will then be formed due to $\phi_i$ exchange which is equivalent to the exchange of a pair of the original bifundamental scalars.

In what follows, we will present in some details the chain of breaking $SU(3)^7 \rightarrow SU(3) \times SU(2)_L \times U(1)_Y$ for one class of fermions, Class A, and simplify the discussion for Class B. The chain of symmetry breaking that we wish to accomplish is one in which one ends up with four chiral generations: $4[(3,\bar3,1)+(1,3,\bar3)+(\bar3,1,3)]_F$. In the breaking of $SU(3)^7 \rightarrow SU(3)^6 \rightarrow SU(3)^5 \rightarrow SU(3)^4 \rightarrow SU(3)^3$, some fermions of the original set will acquire a dynamical mass leaving four massless  chiral generations at the trinification level. This will be shown below.

\begin{itemize}
\item Class A: ~ Consider the fermions $\sum_{i=1}^{7~\textrm{mod}[7]}\,(\,3_i,\,\bar3_{i+1}\,)$. They transform as:
\begin{eqnarray}
(3,\bar3,1,1,1,1,1)+\textrm{perms}
=
 && (3,\bar3,1,1,1,1,1) \nonumber \\
&+& (1,3,\bar3,1,1,1,1) \nonumber \\
&+& (1,1,3,\bar3,1,1,1) \nonumber \\
&+& (1,1,1,3,\bar3,1,1) \nonumber \\
&+& (1,1,1,1,3,\bar3,1) \nonumber \\
&+& (1,1,1,1,1,3,\bar3) \nonumber \\
&+& (\bar3,1,1,1,1,1,3)\,.
\end{eqnarray}

\begin{itemize}

\item  $SU(3)^7 \rightarrow SU(3)^6$:

The relevant fields for the first step of symmetry breaking are
\begin{eqnarray} \label{chi1}
&&\chi_1 \equiv (1,1,3,\bar{3},1,1,1)_F \,, \\
&&\phi_1 \equiv (1,1,3,\bar{3},1,1,1)_S \,,
\end{eqnarray}
where the subscripts $F$ and $S$ stand for fermions and scalars respectively. The Yukawa coupling is
\begin{equation} \label{yukcoupling1}
{\cal L}_{Y1}=g_{\chi_1} \chi_1^{T} C \chi_1 \phi_1 + H.c. \,,
\end{equation}
where $C$ is the  charge conjugation matrix.  Following \cite{Hung2}, for a sufficiently large Yukawa coupling $g_{\chi_1} $, a condensate (a bound state of $\chi_1^{T} C \chi_1$ ) can get formed by the exchange of a massless $\phi_1$:
\begin{equation} \label{cond1}
\langle \chi_1^{T} C \chi_1 \rangle = \langle (1,1,\bar{3},3,1,1,1) \rangle \sim -\Lambda_{C1}^3 \,,
\end{equation}
 where $\Lambda_{C1}$ is the scale where the boundary condition similar to Eq.(\ref{bc1}) is obeyed. The condensate in Eq. (\ref{cond1}) breaks {\em two} of the seven $SU(3)$ down to a diagonal subgroup $SU(3)$. This accomplishes the first step of symmetry breaking  $SU(3)^7 \rightarrow SU(3)^6$. As summarized in the last section, the fermion which participates in the condensate acquires a dynamical mass of order of the condensate scale. Here it is $\chi_1$ and hence $\chi_1$ acquires a dynamical mass of O($\Lambda_{C1}$) and drops out. Under $SU(3)^6$, one has the following remaining massless fermions:
\begin{eqnarray}
(3,\bar3,1,1,1,1,1)+\textrm{perms}
\rightarrow
 && (3,\bar3,1,1,1,1) \nonumber \\
&+& (1,3,\bar3,1,1,1) \nonumber \\
&+& (1,1,3,\bar3,1,1) \nonumber \\
&+& (1,1,1,3,\bar3,1) \nonumber \\
&+& (1,1,1,1,3,\bar3) \nonumber \\
&+& (\bar3,1,1,1,1,3)\,.
\end{eqnarray}

\item  $SU(3)^6 \rightarrow SU(3)^5$ :

For the next step, the relevant fields are $\chi_2 \equiv (1,3,\bar{3},1,1,1)_F$ and $\phi_2 \equiv (1,3,\bar{3},1,1,1)_S$ with a Yukawa coupling:
\begin{equation} \label{yukcoupling2}
{\cal L}_{Y2}=g_{\chi_2} \chi_2^{T} C \chi_2 \phi_2 + H.c. .
\end{equation}
The condensate is now
\begin{equation} \label{cond2}
\langle \chi_2^{T} C \chi_2 \rangle = \langle (1,\bar{3},3,1,1,1) \rangle \sim -\Lambda_{C2}^3 \,,
\end{equation}
giving  $SU(3)^6 \rightarrow SU(3)^5$ and $\chi_2$ drops out, having acquired a mass of order O($\Lambda_{C2}$). This leaves:
\begin{eqnarray}
(3,\bar3,1,1,1,1,1)+\textrm{perms}
\rightarrow
 && (3,\bar3,1,1,1) \nonumber \\
&+& (1,3,\bar3,1,1) \nonumber \\
&+& (1,1,3,\bar3,1) \nonumber \\
&+& (1,1,1,3,\bar3) \nonumber \\
&+& (\bar3,1,1,1,3)\,.
\end{eqnarray}

\item $SU(3)^5 \rightarrow SU(3)^4$ :

With $\chi_3 \equiv (3,\bar{3},1,1,1)_F$, $\phi_3 \equiv (3,\bar{3},1,1,1)_S$ and the condensate $\langle \chi_3^{T} C \chi_3 \rangle = \langle (\bar{3},3,1,1,1) \rangle \sim -\Lambda_{C3}^3$, one obtains  $SU(3)^5 \rightarrow SU(3)^4$ and  $\chi_3$ drops out, having acquired a mass of order O($\Lambda_{C3}$). This leaves:
\begin{eqnarray}
(3,\bar3,1,1,1,1,1)+\textrm{perms}
\rightarrow
&& (3,\bar3,1,1) \nonumber \\
&+& (1,3,\bar3,1) \nonumber \\
&+& (1,1,3,\bar3) \nonumber \\
&+& (\bar3,1,1,3)\,.
\end{eqnarray}

\item $SU(3)^4 \rightarrow SU(3)^3$ :

The last step contains $\chi_4 \equiv (1,1,3,\bar{3})_F$ and $\phi_4 \equiv (1,1,3,\bar{3})_S$. The condensate $\langle \chi_4^{T} C \chi_4 \rangle = \langle (1,1,\bar{3},3) \rangle \sim -\Lambda_{C4}^3$ finally gives the trinification $SU(3)^4 \rightarrow SU(3)^3$  and $\chi_4$ drops out, having acquired a mass of order O($\Lambda_{C4}$).
We are left with one chiral family:
\begin{eqnarray}
(3,\bar3,1,1,1,1,1)+\textrm{perms}
\rightarrow
&& (3,\bar3,1) \nonumber \\
&+& (1,3,\bar3) \nonumber \\
&+& (\bar3,1,3)\,.
\end{eqnarray}

\item $SU(3)^3 \rightarrow SU(3) \times SU(2)_L \times U(1)_Y$ :

Following the steps presented above and using the particle content listed in \eqref{content1}, \eqref{content2} and \eqref{content3}, one can form the following color-singlet bilinears:
\begin{eqnarray}
\label{bilinear}
\psi_L^T C \psi_L& = & (1,\bar{3}+6, 3+\bar{6}) \,, \nonumber \\
\psi_{Q}^T C \psi_{Q^c} &=& (1+8, \bar{3},3) \,.
\end{eqnarray}
Let us recall that the original scalars are in the bifundamental representation 3\,$\sum_{i=1}^{7~\textrm{mod}[7]}\,(\,3_i,\,\bar3_{i\pm 3}\,)$ and transform as $\{\,3\,[(3,1,1,\bar3,1,1,1)+\textrm{perms}]+h.c.\,\}_S\,$. The relevant massless scalars, for our present purpose, which remain at the trinification level are $\phi_{1,2,3} = (1, 3,\bar3)$. One can have  Yukawa couplings of the type
\begin{equation}
 \label{yuktrini}
 {\cal L}_{Y3}=(g_{L}\psi_L^T C \psi_L + g_{Q} \psi_{Q}^T C \psi_{Q^c}) \phi^\dagger_1 +h.c. \,,
 \end{equation}
and similarly for other couplings with $\phi_{2,3}$. Following similar arguments as above, one can obtain condensates of the form $\langle  {\cal N}_{1,2L} ^{T} C  {\cal N}_{1,2L} \rangle$, $\langle  {\cal B}^{c, T} C B_L \rangle$ which breaks $SU(3)^3$ down to $SU(3) \times SU(2)_L \times U(1)_Y$. Details of this breaking will be treated in a longer version.

\end{itemize}

\item Class B: ~ Consider the fermions $3\,\sum_{i=1}^{7~\textrm{mod}[7]}\,(\,3_i,\,\bar3_{i+2}\,)$. For each of the 3 generations, the fermions transform as:
\begin{eqnarray}
(3,1,\bar3,1,1,1,1)+\textrm{perms}
=
&&  (3,1,\bar3,1,1,1,1) \nonumber \\
&+& (1,3,1,\bar3,1,1,1) \nonumber \\
&+& (1,1,3,1,\bar3,1,1) \nonumber \\
&+& (1,1,1,3,1,\bar3,1) \nonumber \\
&+& (1,1,1,1,3,1,\bar3) \nonumber \\
&+& (\bar3,1,1,1,1,3,1) \nonumber \\
&+& (1,\bar3,1,1,1,1,3) \,.
\end{eqnarray}

The breaking $SU(3)^7 \rightarrow SU(3)^6$ is accomplished by the condensate of $\chi_1$ (same as in Class A) as we have seen above. Under $SU(3)^6$, class B reduces to:
\begin{eqnarray}
(3,1,\bar3,1,1,1,1)+\textrm{perms}
\rightarrow
&&  (3,1,\bar3,1,1,1) \nonumber \\
&+& (1,3,\bar3,1,1,1) \nonumber \\
&+& (1,1,3,\bar3,1,1) \nonumber \\
&+& (1,1,3,1,\bar3,1) \nonumber \\
&+& (1,1,1,3,1,\bar3) \nonumber \\
&+& (\bar3,1,1,1,3,1) \nonumber \\
&+& (1,\bar3,1,1,1,3) \,.
\end{eqnarray}

From this set, one can use $\chi_2^{\prime} \equiv (1,3,\bar{3},1,1,1)_F$  to form the condensate  $\langle \chi_2^{\prime,T} C \chi_2^{\prime} \rangle \equiv \langle (1,3,\bar{3},1,1,1)_F \rangle$. As with Class A, one has $SU(3)^6 \rightarrow SU(3)^5$ and the reduction is now:
\begin{eqnarray}
(3,1,\bar3,1,1,1,1)+\textrm{perms}
\rightarrow
&&  (3,\bar3,1,1,1) \nonumber \\
&+& (1,3,\bar3,1,1) \nonumber \\
&+& (1,3,1,\bar3,1) \nonumber \\
&+& (1,1,3,1,\bar3) \nonumber \\
&+& (\bar3,1,1,3,1) \nonumber \\
&+& (1,\bar3,1,1,3) \,.
\end{eqnarray}

The condensate of $\chi_3$ (from Class A) gives $SU(3)^5 \rightarrow SU(3)^4$ and the reduction is now:
\begin{eqnarray}
(3,1,\bar3,1,1,1,1)+\textrm{perms}
\rightarrow
&& (3,\bar3,1,1) \nonumber \\
&+& (3,1,\bar3,1) \nonumber \\
&+& (1,3,1,\bar3) \nonumber \\
&+& (\bar3,1,3,1) \nonumber \\
&+& (\bar3,1,1,3) \,.
\end{eqnarray}

Now $\chi_4^{\prime} \equiv (1,1,3,\bar{3})_F$ and its associated condensate $\langle \chi_4^{T\,\prime} C \chi_4^{\prime} \rangle = \langle (1,1,\bar{3},3) \rangle$ contributes to the final breaking to the trinfication group $SU(3)^4 \rightarrow SU(3)^3$.

There is also a vector-like representation:
$\chi_V \equiv (3,1,\bar{3},1)$, $\chi_V^c \equiv (\bar{3},1,3,1)$ which has a Yukawa coupling to a singlet (at this stage) scalar, $\phi_S \equiv (1,1,1,1)$, which comes from any of the original scalars. The condensate $\langle \chi_V^c \chi_V \rangle$ gives a dynamical mass to $\chi_V $.

Finally, for each of the 3 generations of the fermions, the end result is the reduction:
\begin{eqnarray}
(3,1,\bar3,1,1,1,1)+\textrm{perms}
\rightarrow
&& (3,\bar3,1) \nonumber \\
&+& (1,3,\bar3) \nonumber \\
&+& (\bar3,1,3) \,.
\end{eqnarray}
Therefore, one obtains three chiral families which arises by a different route from the previous single chiral family. The fermions $\chi_2^{\prime} $, $\chi_4^{\prime}$ and $\chi_V$ become massive.


\end{itemize}

In summary, by adding the results for both of the two classes, we have shown how fermion condensates in the $Z_7$ model can break $SU(3)^7$ down to the trinification group and then to $SU(3) \times SU(2)_L \times U(1)_Y$, leaving four chiral families.  The intriguing feature of this model is the fact that three of the four chiral families come from the same initial fermion representation, while the fourth comes from a different initial representation. This appears to single out the fourth generation as somewhat different from the other three.

\section{Discussion and Conclusion}

The limitations of  orbifolded models can be overcome with a bottom up approach if the low energy model has a conformal fixed point. In fact, a postulated existence of a heavy fourth
generation gives rise to a quasi fixed point for the Yukawa couplings at a scale $\Lambda_{FP}$ of O(TeV) \cite{Hung}. Furthermore, the 4th generation Yukawa couplings start to grow as one approaches $\Lambda_{FP}$ from below until they pass a critical coupling required for condensate formation (fermion-antifermion bound-state formation through the exchange of a massless scalar), leading to a dynamical breakdown of the electroweak symmetry \cite{Hung2}. We have left out the final stage of symmetry breaking, namely $SU(3)^3 \rightarrow SU(3)_c \times SU(2)_L \times U(1)_Y$. This interesting topic will be treated in a longer version. This is what we refer to in this paper as the bottom-up approach. It was postulated \cite{Hung,Hung2} that at $\Lambda_{FP}$, SM4 merges into a scale-invariant theory or, better still, into a conformal-invariant theory.

In the top-down approach, one can imagine a conformal model that has all the necessary degrees of freedom to accomplish the various steps of symmetry breaking, leaving in the end process the degrees of freedom of SM4. The matching of the two approaches should occur at around the scale $\Lambda_{FP}$ where the full spectrum of the $SU(3)^n$ gauge theory with bifundamental fermions and scalars should appear. This is one of the distinctive predictions of this type of model for which we will explore the detailed implications elsewhere. But it is sufficient to say that these well defined spectra are in the TeV range and hence accessible by the LHC.

The key point in the search for such conformal models within the framework of orbifolded $AdS_5\otimes S^5/\Gamma$ models is that they must give rise to four chiral families. Within this framework, the most attractive scenario is one which leads to a trinification gauge group $SU(3)^3$ endowed with four chiral families and hence the choice of $N=3$. In general, we search for $Z_n$'s which have sufficient degrees of freedom to accomplish the symmetry breaking $SU(3)^n \rightarrow SU(3)^3$ and provide four chiral families with low scale conformality that avoids gauge coupling degeneracy, allowing the initial value of $\sin ^2 \theta_W^0$ at the conformal breaking scale to lie within a reasonable range so that it may agree with experiment when it gets run down to the scale $M_Z$. Although we have found four models which fit this bill: $Z_7$, $Z_{11}$, $Z_{13}$ and $Z_{14}$, we have only discussed the $Z_7$ model to show how a dynamical symmetry breaking can be accomplished by condensates of fermions in analogy with the scenario of dynamical electroweak symmetry breaking presented in \cite{Hung2}. The other models are also very promising and their theoretical and
phenomenological ramifications will be presented in a longer version of this paper. For a different approach to ultraviolet completion of the standard model, see \cite{Burt}.

\section*{Acknowledgments.}~~
We thank Paul Frampton for a useful discussion. The work of  C.M.H. and T.W.K.
was supported by US DOE grant DE-FG05-85ER40226. The work of PQH was supported by US DOE grant DE-FG02-97ER41027.

\end{document}